# ON A LINK BETWEEN CLASSICAL PHENOMENOLOGICAL LAWS OF GASES AND QUANTUM MECHANICS

Tolga Yarman, Okan University, Akfirat, Istanbul, Turkey (tyarman@gmail.com) &
Savronik, Organize Sanayii Bölgesi, Eskisehir, Turkey, tolgayarman@gmail.com
Alexander L Kholmetskii, Belarus State University, Minsk, Belarus (kholm@bsu.by)
Önder Korfali, Galatasaray University, Ortaköy, Istanbul, Turkey

**ABSTRACT**

In this paper we find a connection between the macroscopic classical laws of gases and the quantum mechanical description of molecules, composing an ideal gas. In such a gas, the motion of each individual molecule can be considered independently on all other molecules, and thus the macroscopic parameters of ideal gas, like pressure *P* and temperature *T*, can be introduced as a result of simple averaging over all individual motions of molecules. It is shown that for an ideal gas enclosed in a macroscopic cubic box of volume *V*, the *constant, in the classical law of adiabatic expansion*, i.e. $PV^{5/3} = const$, can be derived, based on quantum mechanics. Physical implications of the result we disclose are discussed. In any case, our finding proves, seemingly for the first time, a *macroscopic manifestation of a quantum mechanical behavior,* and this in relation to classical thermodynamics.

## 1. Introduction

Time to time, most of us, no doubt just like many scientists of the 20[th] century, were puzzled with the question of finding a bridge between the *Boltzmann constant k* and the *Planck constant h*. In particular, de Broglie already in his doctorate thesis has brilliantly applied his relationship *(associating a wave length with the momentum of a moving particle)* to the statistical equilibrium of gases [1], but did not advance his idea, to see whether one can along such a line obtain anything related to the law of gases, established long ago, in 1650.

The *Boyle-Mariotte law of ideal gas* is given, as usual, by

$$PV = nRT = kNT \qquad (1)$$

with the following designations: *P* is pressure of the gas, *V* volume of the gas, *T* temperature of the gas, n= $N/N_A$ number of *moles* the gas is made of, *N* number of molecules in the gas, $N_A$ the Avogadro number, *R* the gas constant, and $k=R/N_A$ the Boltzmann constant.

The Kinetic Theory of Gases allows us to derive the same casing as that of Eq. (1) via considering the momentum change of molecules when bouncing back from a wall of the container [2]. Assuming for simplicity a cubic geometry, one obtains[1]

---

[1] For the sake of completeness, let us recall the classical derivation of Eq. (2). The force $f_x$ exerted by the molecule of mass *m* and velocity *v*, delineating $v_x$ as its *x*-component, on the wall orthogonal to *x*, is given by Newton's second law, i.e. $f_x = -\Delta p_x / \Delta t$, where $\Delta p_x = -2mv_x$ is the *algebraic increase in the momentum,* whilst the molecule bounces back from the wall, and $\Delta t = 2L/v_x$, *L* being the size of the container along the *x*-direction. Thus, $f_x$ becomes $f_x = mv_x^2 / L$. We can suppose that we deal with "an average" molecule, and all molecules behave as this "average" molecule. Hence, summing over *N* molecules, the gas is made of, we get the total force

$$F_x = N\frac{m\overline{v_x^2}}{L} = \frac{N}{3}\frac{m\overline{v^2}}{L} ,$$

where we have the mean square velocities; recall that at the equilibrium the mean square velocities, for all directions, point to the same quantity. The pressure *P* exerted by *N* molecules on the wall of concern, is thence

$$PV = \frac{N}{3} m\overline{v^2} = \frac{2}{3} N\overline{E}, \qquad (2)$$

$\overline{E} = m\overline{v^2}/2$ being the average translational energy of molecules of mass *m*.

The comparison of this relationship with Eq. (1) yields

$$\overline{E} = \frac{3}{2} kT . \qquad (3)$$

Furthermore, Eq. (2), given the way it is framed (cf. footnote 1), can well be written for the pressure *p*, that would be built in a volume *V*, containing *just one molecule* of *translational energy* $\overline{E}$:

$$pV = kT = \frac{2}{3}\overline{E} . \qquad (4)$$

Could this equation be a basis to build a *bridge* between the law of gases (mainly characterized by the Boltzmann constant), and quantum mechanics (which will evidently involve the energy quantity $\overline{E}$)? Here though, while the equality $pV = kT$ points to the law of gases, the next equality $kT = (2/3)\overline{E}$ is no more than a definition of the temperature, in terms of the average translational energy of the molecules $\overline{E}$. So $kT = (2/3)\overline{E}$ is not to provide us with a bare relationship between *k* and *h*, at all.

Accordingly, Eq. (4) does not bring us anywhere in finding a bridge between the *law of gases* and *quantum mechanics*.

In other terms, $\overline{E}$ is to be expected to involve the Planck constant, yet this, via Eq. (4) does not provide us with a relationship between *h* and *k*, for it would yield merely a relationship between *h* and *kT*. Hence, based on Eq. (4) we are bound to fail to establish a relationship between macroscopic properties of an ideal gas and the quantum mechanical description of its molecules.

Thereby we learn that, when we propose to draw a line between the law of gases and quantum mechanics, we should not really look for a relationship between *h* and *k*. Any such effort will be dissolved through a plain definition of the temperature, in terms of the average translational energy of the molecules, and nothing beyond. However, we can still go ahead to check whether the phenomenological laws of gases are well matched to quantum mechanics, if we could explore those laws of gases, which do not involve the constants *R* or *k*. That is the key point of our approach.

## 2. The compatibility of the law of gases with quantum mechanics, based on the constancy of $PV^\gamma$ for an adiabatic transformation

There is a relationship satisfying the criteria we have just set; this is the one describing an *adiabatic transformation* of gases in a wide temperature range, i.e.

$$PV^\gamma = const, \qquad (5)$$

obtained in the familiar way based on the law of gases, considered together with the first law of thermodynamics [3], with the usual definition

---

$$P = \frac{N}{3} \frac{m\overline{v^2}}{L^3},$$

which is Eq. (1), along with $V=L^3$. Note that the foregoing derivation is well based on the formulation of the pressure exerted by just *one molecule* on the given wall; thus it is surely valid for solely one molecule of ideal gas, in which case *N*=1.

$$\gamma = \frac{C_P}{C_V}, \tag{6}$$

where

$$C_V = \frac{3}{2}R, \quad C_P = \frac{5}{2}R, \tag{7},(8)$$

$C_V$ being the heat to be delivered to one mole of ideal gas at constant volume to increase its temperature as much as 1°K, and $C_P$ being the heat to be delivered to one mole of ideal gas at constant pressure to increase its temperature still as much as 1°K. Eqs. (7) and (8) are *exact*, when internal energy levels of molecules are not excited. Such an assertion is fulfilled for an ideal gas by definition. And we will find out that an *ideal gas*, is in fact, a *gas which is made of non-interacting molecules*, each behaving as a *simple quantum mechanical particle* locked up (potential-wise speaking), in an infinitely high box.

From Eqs. (6)-(8), one has $\gamma = \frac{5}{3}$. Since for an ideal gas, all molecules move independently from each other, Eq. (5) remains valid, even if the gas consists in *just one molecule*. And once again, within the frame of the kinetic theory of gases, one first has to express the pressure for one molecule only, before he proceeds for all molecules, making up the gas, in order to formulate the macroscopic pressure, the gas exerts on the walls of the container (cf. footnote 1).

It should be stressed that Eq. (5), as expected, embodies neither the temperature *T*, nor the average translational energy $\overline{E}$, so we are well off the incorrectly set, *dead end problem* we reported above, regarding the search for a pointless bare link between *k* and *h*.

Further on, we would like to introduce the following fundamental question: *Is Eq.(7) compatible with a corresponding quantum mechanical frame, one would set?*

Let us thus consider a particle of mass *m* with a fixed internal energy state, located in a macroscopic cube of side *L*. The non-relativistic Schrödinger equation furnishes the energy $E_n$ at a given motional level, i.e.

$$E_n = \frac{h^2}{8m}\left(\frac{n_x^2}{L^2} + \frac{n_y^2}{L^2} + \frac{n_z^2}{L^2}\right) = \frac{h^2\left(n_x^2 + n_y^2 + n_z^2\right)}{8mL^2}, \tag{9}$$

where we denoted $n_x = 1,2,3,...$, $n_y = 1,2,3,...$, $n_z = 1,2,3,...$, the quantum numbers to be associated with the corresponding wave function dependencies, on the respective directions *x*, *y* and *z*. For brevity, we introduced the number $n^2 = n_x^2 + n_y^2 + n_z^2$ which denotes the specific state characterized by the set of integer numbers $n_x$, $n_y$, and $n_z$. (These should of course, not be confused with the number of moles n (not Italic), introduced in Eq. (1)).

For an ideal gas the "potential energy" within the box, is null. Thus, we have

$$E_n = \frac{1}{2}mv_n^2, \tag{10}$$

$v_n$ being the velocity of the particle at the $n^{th}$ energy level.

At the given energy level, the pressure $p_n$ exerted by *just one particle* on either wall, becomes (cf. Eqs. (2) and (4))

$$p_n = \frac{mv_n^2}{3L^3} = \frac{2}{3}\frac{E_n}{L^3}. \tag{11}$$

Now, let us calculate the product $p_n V^\gamma$ for one particle:

$$p_n V^\gamma = \frac{2}{3} \frac{\frac{h^2 n^2}{8mL^2}}{L^3} \left(L^3\right)^{5/3} = \frac{h^2 n^2}{12m}. \tag{12}$$

This quantity indeed, turns out to be a constant for a given particle of mass *m* at the given energy level. Recall that the total energy $E_n$ of Eq. (9), ultimately determines the *quantized velocity* $v_n$ of Eq.(10).

When it is question of many particles instead of just one, normally, we deal with particles at different, possible, quantized states. This is most likely, what leads to the Maxwellian distribution of particles with different translational energies in a container, at *a given temperature.* It is on the other hand, this temperature specifies *the average particle.* We can visualize the average particle as a single particle, obeying Eq. (12), thus situated at the $n^{th}$ level, and associate the given temperature with the energy coming into play, along with Eq. (3).

Not to complicate things, let us focus on the *average particle,* and suppose that all others behave the same. Furthermore, all three components of the average velocity in equilibrium are expected to be the same. Thus, we can rewrite Eq. (12) for the macroscopic pressure $P_n$ exerted at the given *average state n*, by one mole of gas, on the walls of the container:[2]

$$P_n V^\gamma = N_A \frac{h^2 n^2}{4m}; \tag{13}$$

Eq. (13) discloses the constant involved by Eq. (5). At the average state *n* (i.e. at the given temperature), the mean square speed of the gas molecules is $\overline{v^2}_n$. The average energy $\overline{E}_n$ is furnished accordingly, via the framework of Eq. (10).

Let us calculate what would *n* be for 1mole of $H_2$, delineating the pressure *P* of $10^5$ Pascal in a volume of 1 m$^3$. Then Eq. (13) yields:

$$n \cong \sqrt{\frac{10^5 \times 12 \times 2 \times 10^{-3} \times 1839 \times 0.9 \cdot 10^{-30}}{3 \times 6.023 \times 10^{23} \times 6.62^2 \times 10^{-68}}} \approx 23 \times 10^8. \tag{14}$$

## 3. Discussion

First we remind that the constancy of $PV^\gamma$ under adiabatic transformation of gases is classically well-known result cited in many books on thermodynamics. However, our aim was not to confirm that $PV^{5/3}$ remains constant through an adiabatic transformation; our aim was *(if one could ever), to calculate this particular constant* (which is classically unknown). Thus, in this article, we aimed to bridge *classical thermodynamics* and *quantum mechanics.* Though, we have determined that, toward that aim, it is in vain, to look for a relationship between Bolzmann and Planck constants. Indeed, a relationship involving both *k* and *h,* such as Eq. (4), is nothing more than a definition of say, the *temperature,* in terms of the *translational energy* of the particle in hand.

So, we had to nail down a relationship, which involves neither *k,* not *h,* to be able to work out our goal. Thus, we came out with the task of working out the constancy of $PV^\gamma$.

---

[2] Rigorously speaking, one must write $P_n V^\gamma = \sum_{i=1}^{N_A} \frac{h^2 \left(n_{ix}^2 + n_{iy}^2 + n_{iz}^2\right)}{12m} = N_A \frac{h^2 \overline{n^2}}{4m}$, along with the definition $\frac{1}{3N_A} \sum_{i=1}^{N_A} \left(n_{ix}^2 + n_{iy}^2 + n_{iz}^2\right) = \overline{n^2}$. Thus it becomes clear that, if all particles bared the same set of quantum numbers, each with equal quantum numbers along all three directions, i.e. $n_x = n_y = n_z = n$, then $n = \sqrt{\overline{n^2}}$.

The value of the constancy of $PV^{\gamma}$ is something totally missed over almost a century. As far as records are concerned, no one seems to have even wondered about the possible value of this constant.

Herein we have calculated this constant, based on quantum mechanics, at last making a bridge between *classical thermodynamics,* and *quantum mechanics.*

Our result further makes that the behavior of an ideal gas is nothing, but a *macroscopic manifestation of quantum mechanics.*

Thus, the constancy of $PV^{\gamma}$ *(thus generally, the frame drawn by the law of gases),* happens to be rooted to quantum mechanics, and seems to be deep. It is that the quantity "mass$\times PV^{\gamma}$" turns out to be a *Lorentz scalar.*

Thereby, we expect this scalar to be somehow nailed to a *Lorentz invariant universal constant;* this constant, more specifically, turns out to be $h^2$ *(the square of the Planck constant).*

Accordingly, for a given mass *m*, the quantity $PV^{\gamma}$ relates to $h^2/m$; this is what we have revealed in this article.

Henceforth i) the constancy of $PV^{\gamma}$ appears to be an extension of *quantum mechanics to macroscopic scales,* but even more essentially, ii) it delineates how the *internal dynamics* displayed by a quantum mechanical particle of a given mass, is organized in conjunction with the *size of space*, and the *dynamics* in question take place in, and this universally, at all scales [4, 5]. Here, we will not go in any further details of this fundamental problem.

Following our approach, it is appealing to write the law of gases, for *just one molecule,* in the following form:

$$p_n V = kT_n \quad ; \tag{15}$$

here $p_n$ is the pressure exerted by one molecule; $T_n$ should, along Eqs. (3), (9), be given by

$$T_n = \frac{1}{3}\left(n_x^2 + n_y^2 + n_z^2\right)T_1 = n^2 T_1 , \tag{16}$$

assuming that at the equilibrium, all three quantum numbers are equal to each other (number *n*); $T_1$ is the temperature characterizing the ground level, for which we have $n_x=n_y=n_z=n=1$. This explains the introducing of the coefficient 1/3, in the first RHS of the above equation.
A comparison of Eqs. (3) and (16) yields:

$$E_1 = \frac{3}{2}kT_1 , \tag{17}$$

where $E_1$ is furnished by Eq. (9), still for $n_x=n_y=n_z=n=1$, i.e.

$$E_1 = \frac{3h^2}{8mL^2} . \tag{18}$$

Thus, similarly to Eq. (16), we can write

$$E_n = \frac{1}{3}\left(n_x^2 + n_y^2 + n_z^2\right)E_1 = n^2 E_1 . \tag{19}$$

Here the second equality is valid, if all three quantum numbers are equal to each other, and equal to *n*, in which case, $E_n$ can be associated with the temperature $T_n$.

Hence Eq. (15) (written for just one molecule) becomes

$$p_n V = kn^2 T_1 = \frac{2}{3} n^2 E_1 . \tag{20}$$

Similarly, for one mole of gas, assumed to be made of $N_A$ molecules, all at the average energy level *n,* thus exhibiting the pressure $P_n$ on the walls of its container, we have

$$P_n V = kN_A n^2 T_1 = Rn^2 T_1 = \frac{2}{3} N_A n^2 E_1 . \tag{21}$$

Thence, Eq. (21) displays the *ground level energy* $E_1$, instead of the Boltzmann constant $k$, regardless the fact that $k$ and $E_1$ bare different dimensions. Let us emphasize that $E_1$ is the energy to be associated with the ground level of just one molecule.

Our formulation further replaces the temperature $T$ with $n^2$, regardless the fact that $T$ and $n^2$ bare different dimensions.

Following Eq. (21), the law of gases, i.e. Eq. (1), can be expressed in the form

$$PV = kN_A T = RT = \frac{2}{3} N_A E_1 n^2 = \tilde{E}_1 n^2, \tag{22}$$

where we have defined $\tilde{E}_1 = \frac{2}{3} N_A E_1$ for the sake of aesthetic, in re semblance with the definition $R = kN_A$.